\documentclass{iau} 
\usepackage{graphicx}

\title[Massive stars: stellar models and stellar yields] 
{Massive stars: stellar models and stellar yields, impact on Galactic Archaeology}

\author[Meynet et al.]   
{Georges Meynet$^1$, Arthur Choplin$^1$, Sylvia Ekstr{\"o}m$^1$,
Cyril Georgy$^1$}

\affiliation{$^1$Department of Astronomy, University of Geneva, \\ CH-1290 Versoix, Switzerland \\ email: {\tt georges.meynet@unige.ch} \\[\affilskip]}

\pubyear{2017}
\volume{334}  
\jname{Rediscovering our Galaxy}
\editors{C. Chiappini, I. Minchev, E. Starkenburg \& M. Valentini, eds.}
\begin{document}

\maketitle

\begin{abstract}
The physics of massive stars depends (at least) on convection, mass loss by stellar winds, rotation, magnetic fields and multiplicity.
We briefly discuss the impact of the first three processes on the stellar yields trying to identify
some guidelines for future works.
\end{abstract}

\firstsection 
\section{Introduction}
Despite the fact that massive stars are quite rare (one estimates 3 stars with an initial mass above 8 M$_\odot$ for every 1000 stars on average), they have a dramatic
impact on their immediate surrounding and at larger scales, as the one of galaxies. 
It consists in important injections into the interstellar medium of radiation, kinetic energy and new synthesized elements. 
In this paper we focus on the impact of various ingredients of stellar models on the stellar yields. In a previous review, we discussed the impacts of these same
ingredients on the evolution of massive stars and on the way they end their nuclear lifetime (\cite[Meynet et al. 2017]{2017arXiv170404616M}).


\section{Convection}

Convection is a key process in the physics of stars. The size of the convective cores determines the quantity of fuel available in a given nuclear burning phase and thus has a deep impact on the lifetimes of stars. It also governs the quantity of mass that is transformed into new elements and thus the stellar yields. During the core hydrogen and helium burning phases, convective zones can be considered as chemically homogeneous because the convective mixing timescale is much shorter
than the nuclear burning timescale. 
Convection, as any turbulent process, remains hard to model and
thus such a basic information as the sizes of the convective cores, the limit of the convective regions in general, is still a matter of investigation (see e.g. \cite[Gabriel et al. 2014]{2014A&A...569A..63G}).
Progresses will come mainly from constraints coming from
asteroseismology that allows, at least in slowly rotating stars, to have constraints on the size of the convective core (see the review by \cite[Noels et al. 2015]{2015IAUS..307..470N} and references therein), and through 3D hydrodynamical simulations (see e.g. \cite[Cristini et al. 2017]{2017MNRAS.471..279C}) that
will hopefully provide new recipes for determining the size of the convective cores in 1D-models. 

As reminded just above, stellar yields depend on the size of the convective cores.
For a given initial mass, the larger the convective core, the larger will be the yields of $\alpha$-elements like for instance
$^{16}$O. If we consider a 20 M$_\odot$, the yield of oxygen may vary from 2 to 3.1 M$_\odot$, depending whether the size of the convective core
is obtained from the Schwarzschild criterion or  from the Schwarzschild criterion with a moderate overshoot that increases the core mass by 18\%.
(Meynet 1990).  Such a difference may appear at first sight very significant, but one has to keep
in mind that in a galactic chemical evolution model, the yields of different initial mass stars of various metallicities have to be considered and thus
to the uncertainties linked to convection will be superposed other uncertainties pertaining to the mass loss rates, rotation for instance. In view of the other uncertainties, such a difference is at the moment not strong enough
to change in a significant way the results of the chemical evolution models.

\section{Mass loss}
Mass loss can occur through various processes during the evolution of massive stars (see the review by \cite[Smith 2014]{2014ARA&A..52..487S}): by line driven winds that operate mainly for hot massive stars at near solar metallicity, through the radiative acceleration due to the continuum radiation when the star luminosity is near the Eddington limit, through mechanical mass loss when the surface is rotating sufficiently fast that at the equator the centrifugal acceleration balances the gravity. Mass loss can also be triggered by the presence of a companion through the process of Roche lobe overflow. Let us concentrate here on the effect of the line driven winds. These winds are efficient only when the opacity of the outer layers are large enough to allow some energy in the radiation field to be transferred to the mass in the form of kinetic energy. This will occur typically at solar metallicity for stars with an initial mass above 30-40 M$_\odot$.

Stellar winds have an impact on the yields mainly through the following process: when stellar winds begin to peel off some inner layers of the star whose chemical composition have been changed by nuclear reactions, convection, and/or some other mixing process as those induced by rotation, then the change in these layers are frozen. Indeed, once ejected from the star, the chemical composition of that mass is fixed. If that mass would have been remained locked into the star, then
further processing can still occur, changing its chemical composition and thus the yields. Thus mass loss changes the yields by removing processed material at an early stage of its transformation. 

\begin{figure}[b]
\begin{center}
 \includegraphics[width=7cm]{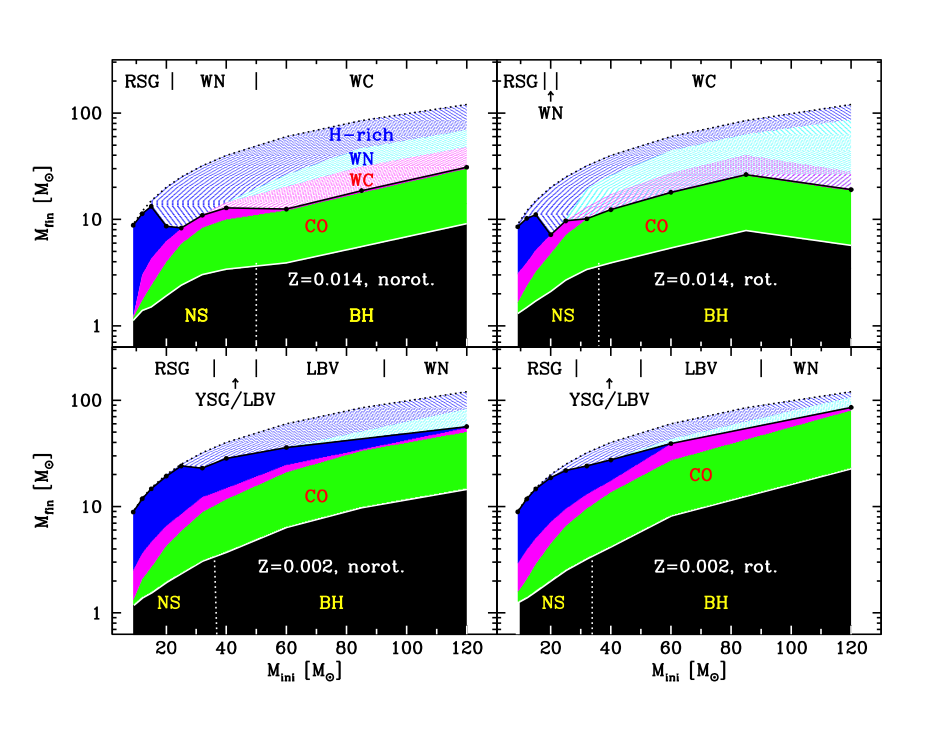}  \includegraphics[width=6.3cm]{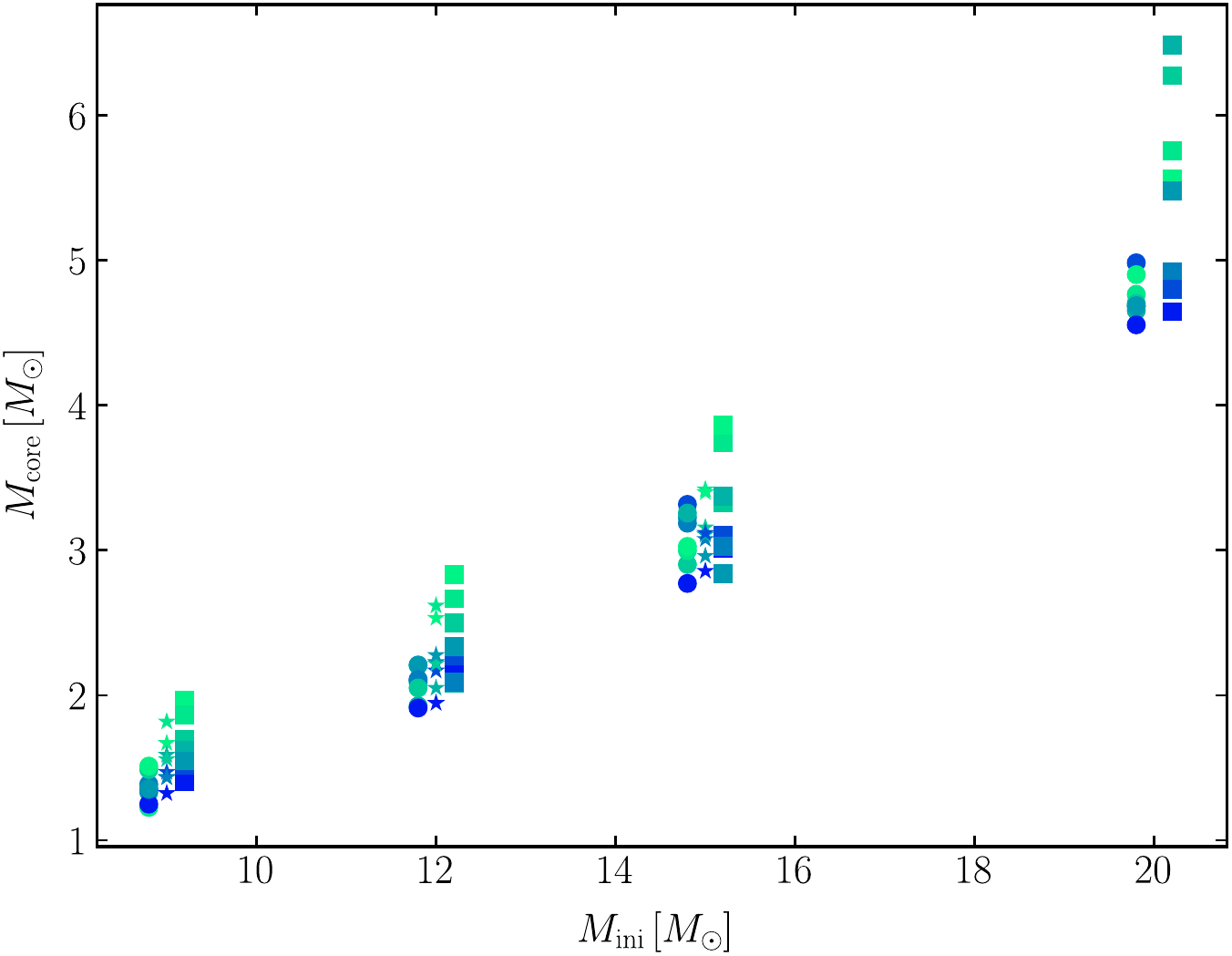}
 \caption{{\it Left: } Masses of the stellar remnants and under the form of supernova and wind ejecta are shown as a function of the initial mass of stars for rotating and non-rotating models at solar metallicity and at the SMC metallicity. The masses that remain locked into the black hole or neutron star correspond to the solid black region. The masses ejected at the time of the supernova,  rich in carbon and oxygen (solid green regions), in H-burning processed material (solid magenta regions),  and ejected with the same abundances as the initial one (solid blue regions) are indicated. The hatched zones are the masses ejected by stellar winds. The blue hatched zone are H-rich stellar winds ejected before the entry of the star into the Wolf-Rayet regime. The cyan and magenta hatched regions correspond to the wind ejected masses when the star is respectively a WN and a WC star. The upper labels indicate the type of the core collapse progenitors. Models from \cite[Ekstr{\"o}m et al. (2010)]{2010A&A...514A..62E} and from \cite[Georgy et al. (2013)]{2013A&A...558A.103G}. Figure taken from \cite[Meynet \& Maeder (2017)]{2017HAND}.
{\it Right:} Mass of the CO core as a function of the initial mass, initial metallicity and rotation. The circles, stars and squares are for metallicities $Z$ equal to  0.014, 0.006 and 0.002 respectively. Colors from dark blue to dark green 
 span the range of initial rotations between a surface angular velocity equal to 10\% the critical one and 95\% the critical one (steps are 0.1, 0.3, 0.5, 0.6, 0.7, 0.8, 0.9, 0.95 times the surface critical angular velocity on the ZAMS). Models from \cite[Georgy et al. (2013)]{2013A&A...553A..24G}.}  
   \label{mass}
\end{center}
\end{figure}

The elements that are changed by this process are those that can make their path to layers that at a given point will be ejected by the winds, thus that can make their path to sufficient outer parts of the star. Elements like those
produced by H- or He-burning regions might be affected by this effect (like helium, nitrogen, carbon, oxygen, fluorine, aluminium 26, ...). 
This point is illustrated in the left panel of Fig.~\ref{mass}.  One sees that at Z=0.014, the mass ejected by stellar winds is much larger than at Z=0.002. One sees also that only at Z=0.014, winds are strong enough for ejecting helium-burning products. This ejected material
will have a chemical composition characteristic of the beginning of the core He-burning phase, namely still very rich in helium and where mainly carbon is synthesized through the triple alpha reaction. Thus winds promote the ejection of carbon
and disfavor the production of oxygen. Indeed mass loss reduces the part of the star that will be processed during the whole core helium burning phase and that ultimately will be transformed mostly in oxygen.
This effect may allow massive stars to compete with the AGB for producing carbon at solar and higher metallicities (\cite[Maeder 1992]{1992A&A...264..105M}).

\begin{figure}[b]
\begin{center}
 \includegraphics[width=5.8cm]{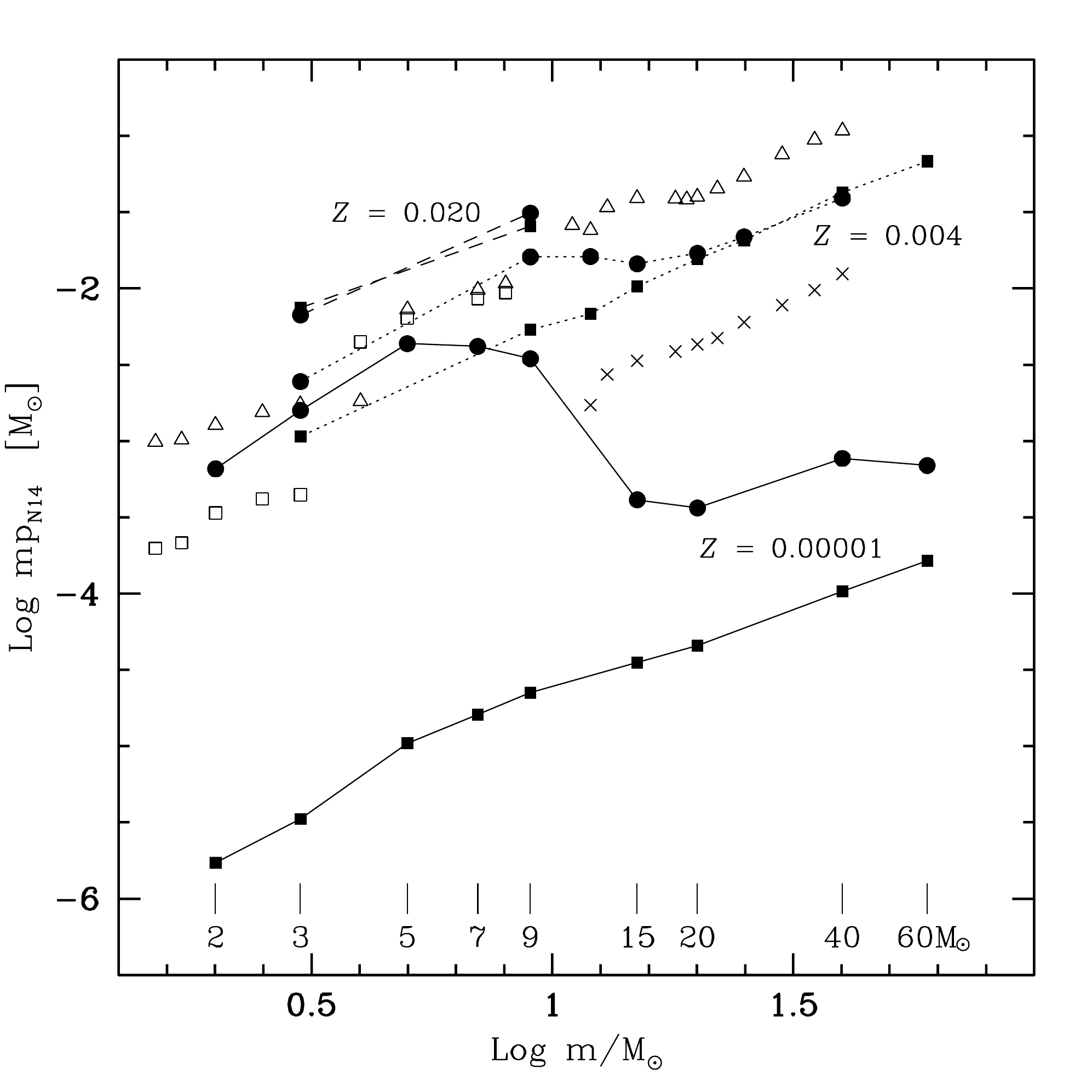} \includegraphics[width=5.6cm]{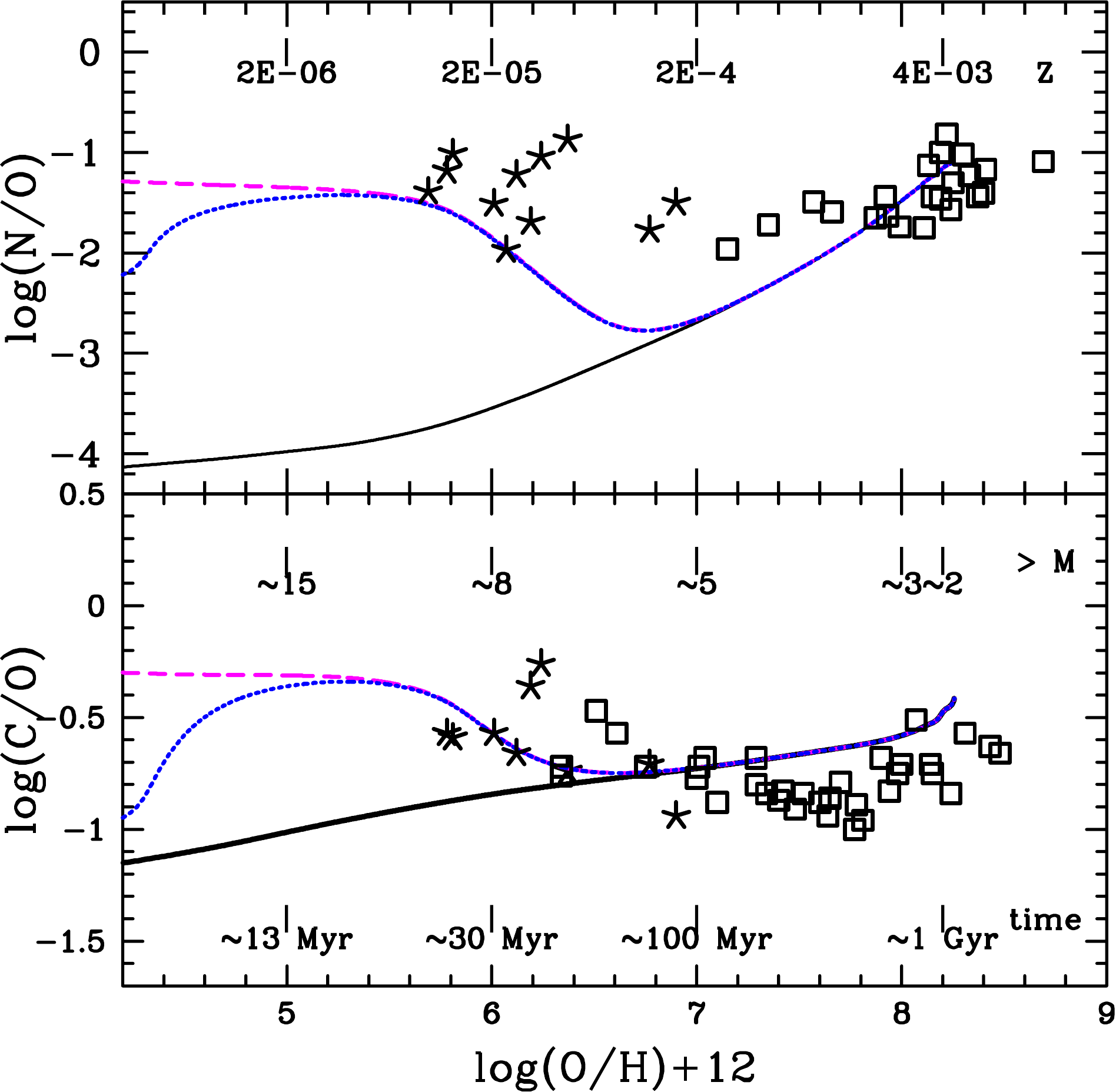} 
 \caption{{\it Left panel} Variation of the stellar yields in $^{14}$N as a function of the initial mass for different
metallicities and rotational velocities. The continuous lines refer to the models at $Z =10^{-5}$ of \cite[Meynet \& Maeder (2002)]{2002A&A...390..561M},
the dotted lines show the yields from the models at $Z$ = 0.004 from \cite[Maeder \& Meynet (2001)]{2001A&A...373..555M}, the dashed lines
present the yields for two solar metallicity models (\cite[Meynet \& Maeder 2002]{2002A&A...390..561M}). The filled squares and circles indicate the cases
without and with rotation respectively. In this last case $v_{\rm ini}$ = 300 km s$^{-1}$. The crosses are for the models of 
\cite[Woosley \& Weaver (1995)]{1995ApJS..101..181W} at $Z=0.1{\rm Z}_\odot$, the empty squares for the yields from \cite[van den Hoek \& Groenewegen (1997)]{1997A&AS..123..305V} at $Z$=0.004, the empty triangles are for solar metallicity models of van \cite[van den Hoek \& Groenewegen (1997)]{1997A&AS..123..305V} up to 8 M$_\odot$ and of \cite[Woosley \& Weaver (1995)]{1995ApJS..101..181W}  above. Figure taken from \cite[Meynet \& Maeder (2002)]{2002A&A...390..561M}. {\it Right panel:}
Evolution of the N/O and C/O ratios.
The (black) solid curve is the chemical evolution model (\cite[Chiappini et al. 2006]{2006A&A...449L..27C}) obtained with the stellar yields of slow rotating $Z=10^{-5}$ models from \cite[Meynet \& Maeder (2002)]{2002A&A...390..561M} and \cite[Hirschi et al. (2004)]{2004A&A...425..649H}. The (magenta) dashed line includes the yields of fast rotating $Z=10^{-8}$ models from \cite[Hirschi (2007)]{2007A&A...461..571H}  at very low metallicity. The (blue) dotted curve is obtained using the yields of the $Z=0$ models presented in 
\cite[Ekstr{\"o}m et al. (2008)]{2008A&A...489..685E} up to $Z=10^{-10}$.  Open squares are observations from \cite[Israelian et al. (2004)]{2004A&A...421..649I}  and stars from  \cite[Spite et al. (2005)]{2005A&A...430..655S}. Figure from \cite[Ekstr{\"o}m et al. (2008)]{2008A&A...489..685E}.}
   \label{nprim}
\end{center}
\end{figure} 

\begin{figure}[b]
\begin{center}
 \includegraphics[width=6.4cm]{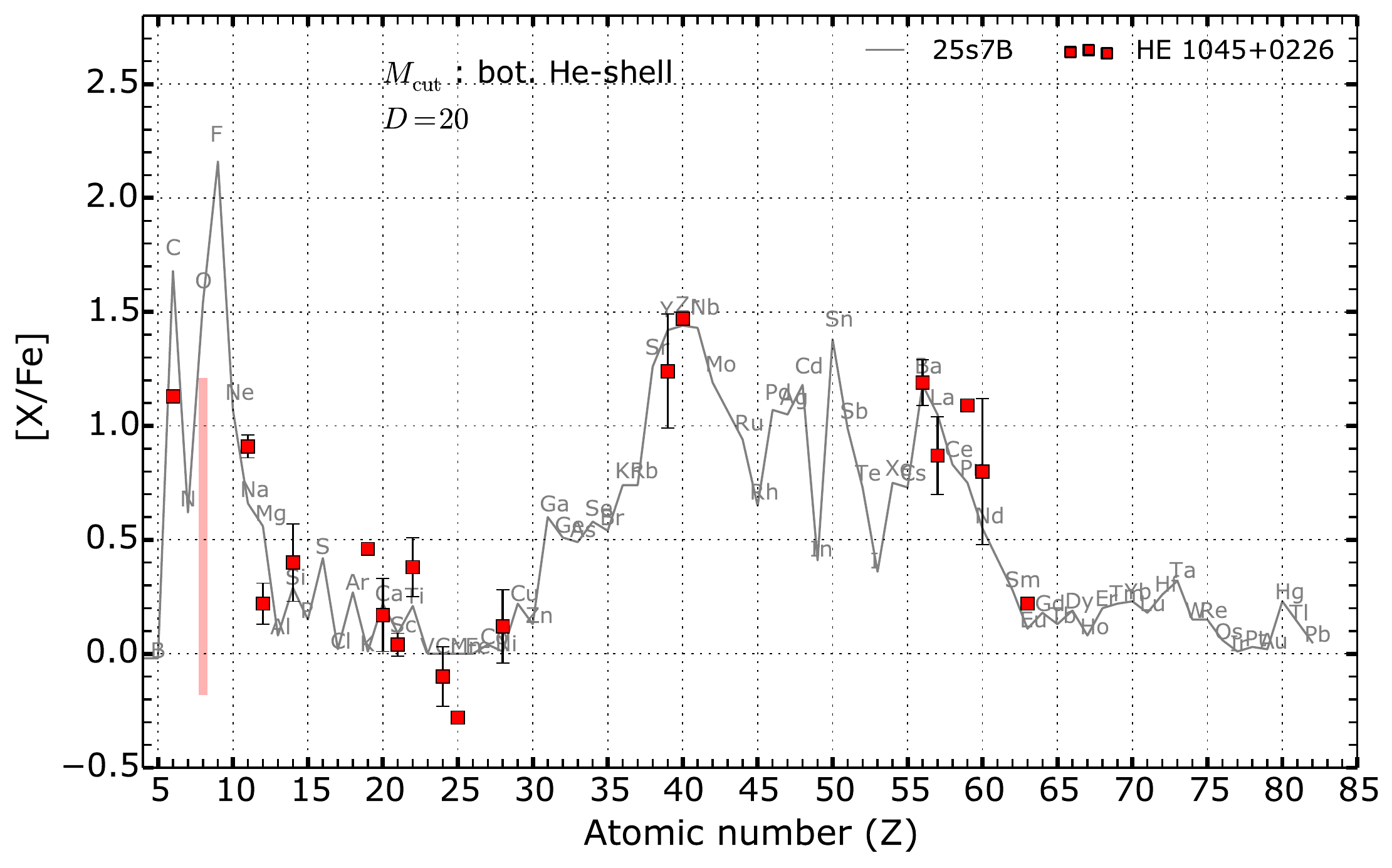}\includegraphics[width=6.4cm]{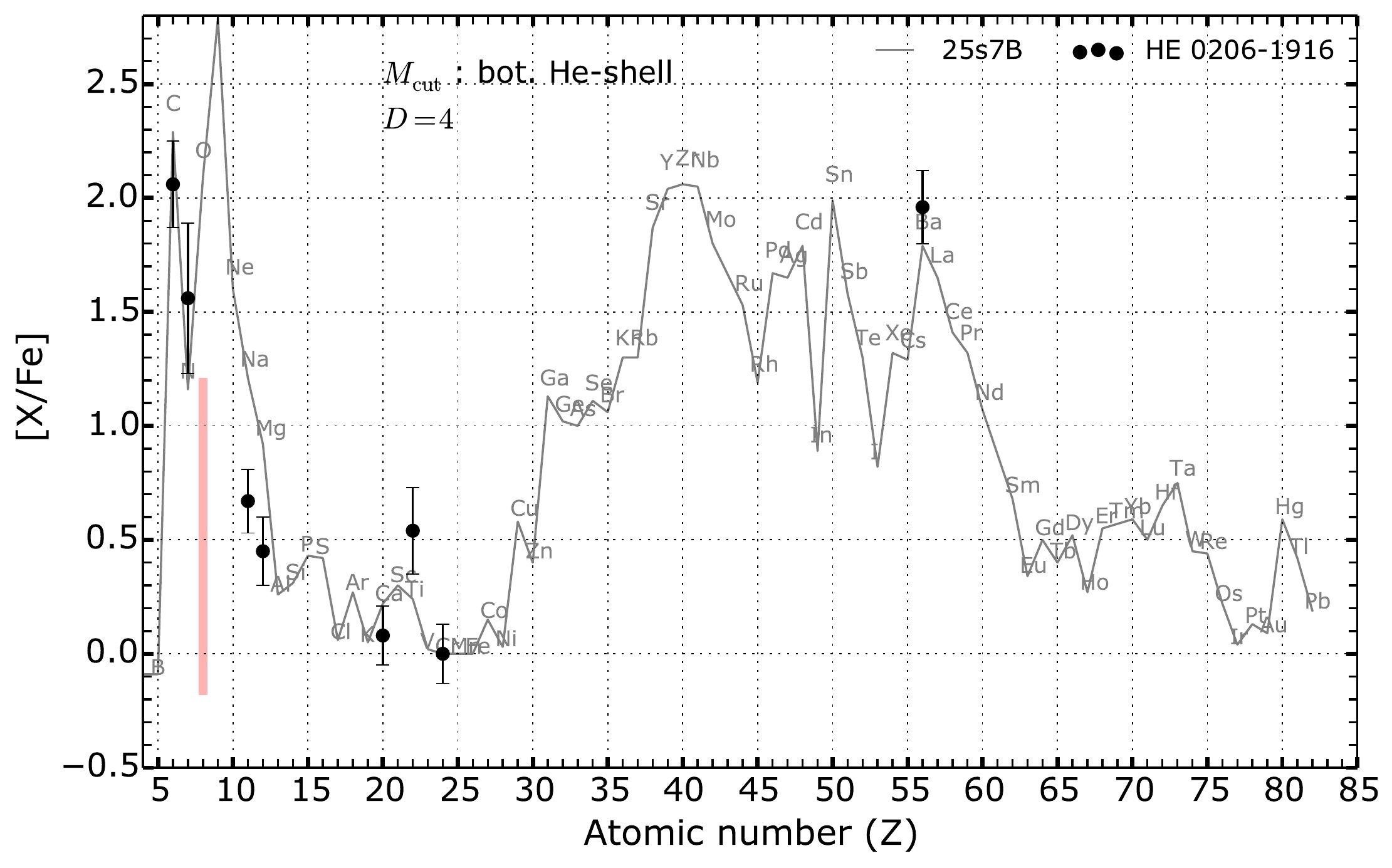}
 \caption{Comparison of the material ejected by a $Z$=10$^{-3}$ 25 M$_\odot$ model with an initial equatorial velocity equal to 70\% the critical velocity (or 490 km s$^{-1}$, solid patterns) with the chemical composition of apparently single CEMP-s stars, {\it left: } HE1045+0226 and {\it right: } HE 2330-0555 (\cite[Hansen et al. 2016]{2016A&A...588A...3H}). 
 The ejecta of the source star is made of wind plus supernova with a mass cut set at the bottom of the He-shell. 
 The dilution factor $D=M_{\rm ISM}/M_{\rm ej}$ is indicated. 
 The red vertical band at $Z=8$ shows the range of [O/Fe] ratios predicted by the AGB models of \cite[Karakas (2010)]{2010MNRAS.403.1413K}. 
 These models have $1<M_{\rm ini}<6$ $M_{\odot}$ and metallicities of $Z=0.004$ and $Z=0.0001$. Figure taken from \cite[Choplin et al. (2017)]{2017arXiv171005564C}.}
  \label{cemps}
\end{center}
\end{figure}

\section{Rotation}
Angular momentum is everywhere in the Universe and in general there is always too much angular momentum. For instance, angular momentum has to be removed at the moment of formation of star in order to explain
their observed rotation rate (\cite[Haemmerl{\'e} et al. 2017]{2017A&A...602A..17H}) . Massive stars, with average surface velocities between 100 and 200 km s$^{-1}$, are in general much faster rotators than the Sun (2 km s$^{-1}$). Rotation induces many instabilities in stellar interiors, instabilities that in turn modify the angular momentum distribution as well as the distribution of the chemical elements (see the book by \cite[Maeder 2009]{2009pfer.book.....M} for a detailed description of the physical processes induced by rotation in stars). 

As we shall see below, rotation may change the yield of some elements by very large factors, up to two orders of magnitudes and even more. This occurs at very low metallicities and is thus particularly important for the nucleosynthesis of the first generations of massive stars in the Universe. Before to describe how rotation deeply affects the nucleosynthesis of metal poor massive stars, it is important to remind that the physics used at low metallicity is the same
as the one used at high metallicity to reproduce observed characteristics of massive stars. For instance, OB Main-Sequence (MS) stars show nitrogen enrichments at their surface that can be well explained by rotating models (\cite[Martins et al 2015]{2015A&A...575A..34M}). The observation of boron depletion at the beginning of the MS of massive stars, not accompanied by a nitrogen enrichment, supports the view of a progressive mixing process that is well compatible with rotational mixing
(\cite[Fliegner et al. 1996]{1996A&A...308L..13F}; \cite[Frischknecht et al. 2010]{2010A&A...522A..39F}; \cite[Proffitt et al. 2016]{2016ApJ...824....3P}). Also the existence of a few Wolf-Rayet stars presenting a surface composition made up of some mixture of both H and He-burning products (the so-called WN/C stars) supports some internal mixing process (see e.g. the discussion in \cite[Meynet \& Maeder 2003]{2003A&A...404..975M}).

Rotation appears a reasonable engine for driving an internal mixing. Indeed it is a well known fact that many instabilities like shear instabilities or meridional currents are triggered by rotation. 
Since rotation implies some turbulent processes and since turbulence cannot be fully described in 1D numerical models, some parameters have to be adjusted. In the Geneva model, there are two parameters that have to be adjusted. One of them
is the ratio between the energy that is used to mix the matter and the energy available in the shear. The other parameter describes the interaction between the horizontal shear turbulence and the meridional currents.
These parameter are fixed so that observed surface abundances in typically 10-20 M$_\odot$ solar metallicity models can be reproduced from rotating models showing surface rotation well in the range of observed values (see Fig.~11 in \cite[Ekstr{\"o}m et al.  2012]{2012A&A...537A.146E}). Once these two parameters are fixed, the same physics is used to compute lower or larger initial masses, and for computing models at other metallicities.

In the right panel of figure~\ref{mass}, the CO core masses at the end of the core helium burning phase are shown as a function of the initial mass for three metallicities and nine different initial rotations.
We see  that the CO core is in general increasing when the initial rotation increases. Thus from a qualitative point of view, rotation has a similar effect on the size of the core as an overshoot.
The dispersion of the values  due to rotation is the largest  for the 20 M$_\odot$ model at Z=0.002, even larger than the one expected from changing the overshoot between 0 and 0.2. 
Rotation produces many other effects as shown below, that an overshoot cannot achieve. 
 
At low metallicities, rotating models predict that some mixing occurs between the helium-burning core and the H-burning shell.
Some elements produced in the helium-burning core, as carbon and oxygen migrate into the
H-burning shell. There, these elements are transformed, at least partly, into nitrogen through the CNO cycle. 
One speaks in that case of {\it primary} nitrogen, since the nitrogen produced in that way is produced from a chain of reactions linking hydrogen and helium to nitrogen occurring in the star itself. One says that nitrogen is {\it secondary} when it is produced
from carbon and oxygen initially present in the cloud that formed the star and not from carbon and oxygen synthesized by the star itself. Thus rotation at low metallicity can be responsible for large amounts of primary nitrogen. Other primary isotopes are produced like for instance $^{13}$C.

The nitrogen produced in the H-burning shell can in turn migrate into the helium burning core. When it is engulfed in the He-core, it is transformed into {\it primary} $^{22}$Ne.
This may have dramatic consequences for the production of s-process elements in massive stars.  Indeed the neutrons come from the reaction $^{22}$Ne($\alpha$, n)$^{25}$Mg.
 The more nitrogen produced in the H-burning shell, the more it will be engulfed into the helium core, and the more $^{22}$Ne and neutrons are produced. 
 Now in order to have some s-process elements, it is also needed to have some seeds that can capture these neutrons in order to form heavy elements beyond the iron peak. The main seed
 is iron. Thus some iron is  also needed. While primary nitrogen can be produced from Pop III up to a metallicity around 0.001 in massive rotating stars, s-process elements can be boosted only
in the higher part of the range indicated above, from Z=10$^{-5}$ up to 10$^{-3}$. Below Z=10$^{-5}$, there is not enough iron seeds, above 10$^{-3}$, rotational mixing is not efficient enough (\cite[Pignatari et al. 2008]{2008ApJ...687L..95P}; \cite[Frischknecht et al. 2016]{2016MNRAS.456.1803F}). 
 
 Why is low metallicity required for obtaining substantial amounts of primary nitrogen? The main reason is because at low metallicities stars are more compact and the gradients of the angular velocity are in general steeper. These two effects
 allow for more efficient mixing. More details about the physics of this process can be found in \cite[Maeder \& Meynet (2001)]{2001A&A...373..555M}.  The left panel of Fig.~\ref{nprim}  show yields of nitrogen from various models.
 The lines connecting filled symbols compare Geneva models with and without rotation at three metallicities. We can see that large differences arise between non-rotating and rotating models at very low metallicities (here Z=0.00001).
 The empty symbols are yields from non-rotating models of other authors for metallicities equal and above 0.1 Z$_\odot$.  One sees that  the non-rotating Geneva yields are well in line with these non-rotating yields of author authors
 accounting for the slightly different metallicities considered (of course the 0.001 yields  of \cite[Woosley \& Weaver(1995)]{1995ApJS..101..181W} have to be compared with the yields at Z=0.004 of the Geneva models).  We see that the impact of rotation on the yields of nitrogen may amount to more than 2 orders of magnitude. In that case, rotation has a deep and significant impact that cannot be blurred by any other different effects as was the case for the impact of overshooting.
These are predictions of models and not ad hoc models built in order to produce primary nitrogen at low Z.  Indeed, primary nitrogen production comes out from models that initially were constructed to fit observed surface enrichments of solar metallicity massive stars. Very interestingly, when using yields from rotating models, assuming that most stars will have an initial angular momentum content  of the same order as their solar metallicity counterparts and using these yields in a chemical evolution model for the halo of our Galaxy, a good match can be achieved both for the evolution of the N/O and C/O ratios (\cite[Chiappini et al. 2003]{2003A&A...410..257C}; \cite[Chiappini et al. 2006]{2006A&A...449L..27C}; \cite[Chiappini et al. 2008]{2008A&A...479L...9C}). Let us note that the blue curve in the upper right panel goes below the observed points mainly because in that model the effects
of rotation in models for intermediate mass stars is not accounted for.

The impact of rotating models on the s-process nucleosynthesis in the halo and in the bulge has been discussed in many recent works (\cite[Chiappini et al. 2011]{2011Natur.472..454C}; \cite[Cescutti et al.(2013)]{2013A&A...553A..51C}; \cite[Barbuy et al. 2014]{2014A&A...570A..76B}; \cite[Siqueira-Mello et al. 2016]{2016A&A...593A..79S}). Let us briefly discuss the impact on the Carbon-Enhanced Metal Poor stars (CEMP, see e.g. \cite[Norris et al. 2013]{2013ApJ...762...28N} and references therein).
These stars can be classified in different families depending on their content in neutron-capture elements: some of them present no significant enhancements of these elements (the CEMP-no), other present
significant enhancements in s or r or in s and r process elements (CEMP-s, CEMP-r, CEMPR-r/s). The CEMP-s are explained mainly by accretion of some material coming from an AGB companion (see \cite[Lucatello et al. 2005]{2005ApJ...625..825L}). This model is supported by the very high fraction of binaries among the CEMP-s stars. Recently however, \cite[Hansen et al. (2016)]{2016A&A...588A...3H} have discovered some CEMP-s stars with no detected radial velocity variations. These stars might be isolated and if it is the case, the classic model cannot work. \cite[Choplin et al.(2017)]{2017arXiv171005564C} have investigated to which extend these solitary CEMP-s stars might be formed from the ejected envelope of a metal poor massive rotating star diluted with some interstellar medium material. Among the four solitary CEMP's stars, three of them can indeed be well reproduced by such a model. The cases of HE1045+0226 and HE 0206-1916 are shown in Fig.~\ref{cemps}. The prediction of the model fits reasonably well the observed pattern of the abundances supporting the view that such stars might be formed from the
envelope of massive rotating star at low Z.

Interestingly, the normal halo stars and part of the peculiar CEMP stars (likely the CEMP-no and some CEMP-s) might be formed from the same type of stars, namely massive rotating stars at low metallicity. The normal halo stars come from a relatively well mixed reservoir that has mixed and accumulated the yields of stars of various initial masses, rotations, initial metallicities, the CEMP, in contrast might owe their peculiar abundances from a few, may be only one event, but the common point in this
approach is that, in both case, rotation plays a key role.

To end this brief discussion about the impact of rotation, we can mention two interesting future lines of investigation (among many others that will come out in the future). First concerning the CEMP stars, it has been argued that
lithium can be used to constrain the degree of dilution between the massive star ejecta and the ISM. Indeed one expects that ISM has the cosmological lithium abundance, while the massive star ejecta are completely free from lithium.
The final abundance of lithium in the CEMP stars will thus depend on the respective fraction of massive star ejecta and ISM entering into their composition.
The exercise has been done in \cite[Meynet et al. (2010)]{2010A&A...521A..30M}. It happens that some CEMP stars might be formed from nearly pure massive star ejecta. If correct, these CEMP stars should be helium-rich.
Let us suppose that the abundance of helium might be obtained through observations then this abundance would provide many interesting clues.
In case helium enrichments would be found then it would be an indication of a peculiar mode of star formation from nearly pure ejecta. It would also provide examples of stars that would reflect the most directly the composition
of the envelope of a massive star in the very early Universe. On the contrary, if no helium enhancements is found, this would indicate that likely lithium has been destroyed in the star itself, providing constraints on the physics
of low-mass stars at very low metallicity. A possibility to check whether a CEMP star is helium rich or not would be to find such stars forming an eclipsing binary. In that case the masses of the stars can be determined from the
dynamic of the system, their position in the HR diagram from spectroscopy.  Comparisons with stellar models of various initial helium content would constrain their helium content.

Another interesting question is to know which kind of final fate will meet the stars that produce important amounts of primary nitrogen. As indicated above, these stars should have been rotating. Likely these stars will keep important amount of angular momentum in their core at the moment of the core collapse. Such rapidly rotating cores may produce aspherical supernovae or jet-induced supernova (\cite[Tominaga et al. 2007]{2007ApJ...657L..77T}). Interestingly when low energy deposition rate by the jets is considered,
jet-induced supernovae will eject very iron-poor carbon-rich material similar to the spherical models for faint and fallback supernovae proposed by \cite[Umeda \& Nomoto (2003)]{2003Natur.422..871U}.
Differences in the yield of some elements are expected between these two types of supernovae. 
Typically the jet-induced SNe are able to inject more zinc and cobalt than the faint supernovae with fallback. 
In case the fast rotation needed  for nitrogen production gives birth to jet-induced supernovae, one would thus expect some correlation between the nitrogen abundance and the abundances of zinc and cobalt observed in CEMP-no stars. 
At the moment many zinc and cobalt abundances are only upper limits, preventing to draw firm conclusions. On the other hand,
present data cannot discard this idea.

\section{Conclusion}

In the above discussion, we have briefly discussed the impact of convection, mass loss and rotation on the yields of some elements. Other factors as magnetic fields, multiplicity can also have important effects. Magnetic fields
may quench the mass loss by stellar winds allowing the star to keep more mass until the end of its evolution even at high metallicity allowing to produce massive black holes (\cite[Petit et al. 2017]{2017MNRAS.466.1052P}) or to produce pair instability supernovae (\cite[Georgy et al. 2017]{2017A&A...599L...5G}).
Multiplicity may have a dramatic impact on the evolution of a star allowing it to lose and/or accrete material at different evolutionary phases and allowing also some exchanges of angular momentum between the stellar spin and
the orbital spin. Very interestingly, the first association of a short gamma ray burst with the merging of two neutron stars (\cite[Abbott et al. 2017]{2017ApJ...848L..12A}) might give further insights into the origin of the r-process elements. In cases all the r-process would come from such events, one can wonder which kind of constraints can be obtained on the systems of two neutron stars that produced r-elements observed in very metal-poor stars. 
While the lifetimes for producing the two neutron stars would be around of
ten millions years, still some time is needed for the merging to occur. For producing r-process in the very early universe some short merging timescale system is needed. This is likely an interesting line of research to be explored.

\end{document}